\documentclass{article}

\usepackage{graphicx}
\usepackage{amsmath}

\usepackage{algorithm,amsfonts,amssymb,amsthm}
\usepackage{mathrsfs}
\usepackage{appendix}
\usepackage{url}

\usepackage{cleveref} 
\creflabelformat{equation}{#2(#1)#3}
\crefname{equation}{}{}

\crefname{figure}{Fig.}{}
\crefname{table}{Table}{}
\crefname{section}{Section}{}

\usepackage{ascmac}

\makeatletter
\newcommand{\subsubsubsection}{\@startsection{paragraph}{4}{\z@}%
  {1.0\Cvs \@plus.5\Cdp \@minus.2\Cdp}%
  {.1\Cvs \@plus.3\Cdp}%
  {\reset@font\sffamily\normalsize}
}
\makeatother
\DeclareMathOperator*{\argmin}{argmin}

\usepackage{authblk}
\title{Spheroidal Ambisonics: a Spatial Audio Framework Using Spheroidal Bases}
\author[1]{Shoken Kaneko}
\affil[1]{Department of Computer Science, University of Maryland, College Park, MD 20742, USA}
\date{\today} 
\begin{document}

\maketitle

\begin{abstract}
    Ambisonics is an established framework to capture, process, and reproduce spatial sound fields based on its spherical harmonics representation~\cite{gerzon1973periphony, daniel2003further}.
    We propose a generalization of conventional spherical ambisonics to the spheroidal coordinate system and spheroidal microphone arrays, which represent sound fields by means of spheroidal wave functions.
    This framework is referred to as \emph{spheroidal ambisonics} and a formulation for the case of prolate spheroidal coordinates is presented.
    Spheroidal ambisonics allows analytical encoding of sound fields using spheroidal microphone arrays.
    In addition, an analytical conversion formula from spheroidal ambisonics to spherical ambisonics is derived in order to ensure compatibility with the existing ecosystem of spherical ambisonics.
    Numerical experiments are performed to verify spheroidal ambisonic encoding and transcoding when used for spatial sound field recording.
    It is found that the sound field reconstructed from the transcoded coefficients has a zone of accurate reconstruction which is prolonged towards the long axis of a prolate spheroidal microphone array.
\end{abstract}

\section{Introduction}

Immersive multimedia technologies such as augmented reality (AR) and virtual reality (VR) are receiving much attention recently.
Audio is an indispensable factor in such modes of multimedia, and it is essential to be able to capture, process, and render spatial sound fields with high precision for presentation of plausible AR/VR and the creation of immersive experiences.
The spatial audio framework of ambisonics~\cite{gerzon1973periphony} as well as higher-order ambisonics (HOA)~\cite{daniel2003further} is receiving much attention due to the popularization of AR/VR devices as well as the ability to stream this representation using standard platforms~\cite{youtube, facebook}, and its high compatibility with first-person view AR/VR.
Ambisonic spatial audio capturing and processing consists of a microphone array and designated signal processing algorithms that are used to encode the raw microphone array signal to the spherical harmonics-domain spatial description format, which is referred to the ambisonic signal.
This ambisonic signal is decoded to the signal which is fed to loudspeaker arrays to render the spatial sound field.
Such loudspeaker arrays can also be virtualized by means of binaural technologies \cite{noisternig20033d, zotkin2004rendering, kaneko2016ear} and played back using headphones.
Hence the high compatibility of ambisonics with AR/VR applications that usually resort to binaural transducers for audio playback.

\begin{figure}[htbp]
\centering
\includegraphics[width=12cm]{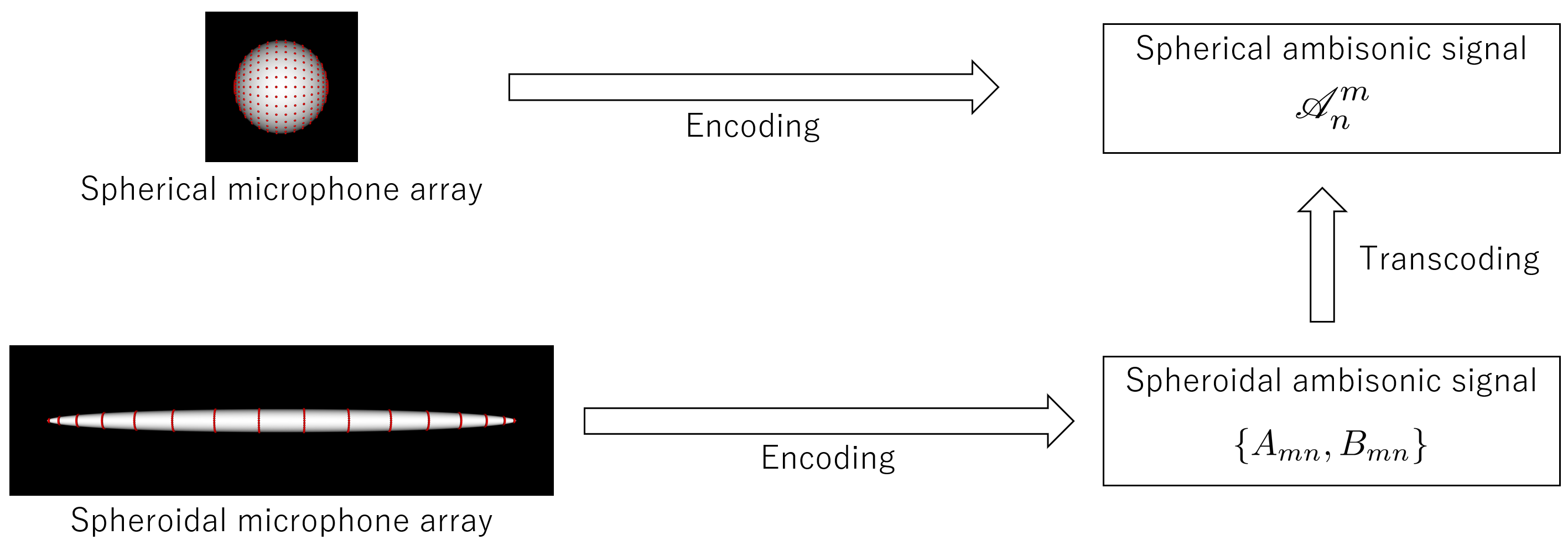}
\caption{The overview of the proposed spheroidal ambisonics. The microphone capsule positions used in the numerical experiments presented in this paper are shown as the red dots in the images.}
\label{fig:micArys}
\end{figure}

Due to its formulation in the spherical harmonics-domain, the most natural implementation of ambisonic recording devices is employing spherical microphone arrays~\cite{gerzon1973periphony,daniel2003further,kaneko2018development}.
In this work, we generalize the framework of ambisonics into spheroidal coordinates and define \emph{spheroidal ambisonics}, which uses spheroidal wave functions for the representation of spatial sound fields.
A formulation for the case of prolate spheroidal coordinates is presented, allowing the use of prolate spheroidal microphone arrays in an analytical manner in contrast to a recently proposed approach which allows arbitrary shaped microphone arrays but relies on numerical simulation to encode the captured field~\cite{zotkin2017incident}.
In addition, an analytical conversion formula from spheroidal ambisonics to spherical ambisonics is derived.
This conversion ability is important to utilize the existing ecosystem around spherical ambisonics after recording the spatial audio with a spheroidal microphone array. 
The overview of the proposed schemes of spheroidal ambisonic encoding and transcoding is shown in \cref{fig:micArys}.
Numerical experiments are performed to validate and demonstrate spheroidal ambisonic encoding and transcoding when used for spatial sound field recording.

\newcommand{\C}{{\mathbb{C}}}
\newcommand{\F}{{\mathbb{F}}}
\newcommand{\R}{{\mathbb{R}}}
\newcommand{\Z}{{\mathbb{Z}}}
\newcommand{\N}{{\mathbb{N}}}
\newcommand{\A}{{\mathbf{A}}}
\newcommand{\B}{{\mathbf{B}}}

\newcommand{\VEC}[2]{\ensuremath{\left(\begin{array}{c} #1  \\ #2 \end{array}\right)}}
\newcommand{\MAT}[4]{\ensuremath{\left(\begin{array}{cc} #1  &  #2 \\ #3 & #4 \end{array}\right)}}

\newcommand{\SA}{{S^{(A)}}}
\newcommand{\SB}{{S^{(B)}}}
\newcommand{\RA}{{R^{(A)}}}
\newcommand{\RB}{{R^{(B)}}}

\section{Background: spherical ambisonics}

The conventional framework of ambisonics, which is referred to as spherical ambisonics, or, simply ambisonics, is briefly reviewed here.
Ambisonic encoding and decoding can be performed by either relying on solving a linear system using least squares~\cite{daniel2003further} or relying on spherical harmonic transformation using numerical integration~\cite{poletti2005three}.
Since the first approach allows more flexibility of the microphone array configuration, this approach is adopted in this paper.
Throughout this paper, only microphone arrays mounted on surfaces of rigid scattering bodies are considered. This is a commonly used approach to avoid the instability arising in encoding filters for hollow microphone arrays due to singularities originating from the roots of the spherical Bessel function~\cite{daniel2003further}.
In this paper, all formulations are presented in the frequency-domain, which can be converted into a time-domain representations by inverse Fourier transform, if necessary.

The spherical harmonics used in this paper are defined as the following.
\begin{equation}
Y _ { n } ^ { m } ( \theta , \varphi ) \equiv \sqrt { \frac { ( 2 n + 1 ) } { 4 \pi } \frac { ( n - m ) ! } { ( n + m ) ! } } P _ { n } ^ { m } ( \cos \theta ) e ^ { i m \varphi },
\end{equation}
with $\theta$ and $\varphi$ the polar and azimuthal angle, respectively, 
and $P_n^m(x)$ the associated Legendre polynomials:
\begin{equation}
P _ { n } ^ { m } ( x ) \equiv ( - 1 ) ^ { m } \left( 1 - x ^ { 2 } \right) ^ { m / 2 } \frac { d ^ { m } } { d x ^ { m } } \left( P _ { n } ( x ) \right),
\end{equation}
with the Legendre polynomials:
\begin{equation}
P _ { n } ( x ) \equiv \frac { 1 } { 2 ^ { n } n ! } \frac { d ^ { n } } { d x ^ { n } } \left( x ^ { 2 } - 1 \right) ^ { n }.
\end{equation}

The above definition of spherical harmonics provides an orthonormal basis:
\begin{equation}
\int _ { \theta = 0 } ^ { \pi } \int _ { \varphi = 0 } ^ { 2 \pi } Y _ { n } ^ { m }( \theta , \varphi ) Y _ { n ^ { \prime } } ^ { m ^ { \prime }  }( \theta , \varphi )^{*} d \Omega = \delta _ { n n ^ { \prime } } \delta _ { m m ^ { \prime } },
\end{equation}
with $\delta _ { ij }$ the Kronecker delta.

\subsection{Encoding in spherical ambisonics}

\def \LambdaE{\Lambda^{\circ}}
\def \LambdaB{\Lambda^{\bullet}}
\def \LambdaBH{\Lambda^{\bullet^{H}}}
\def \LambdaP{\Lambda^{\mathrm{(P)}} } 
\def \LambdaPH{\Lambda^{\mathrm{(P)}^{H}}}
\def \pin{p_{\mathrm{in}}}
\def \ptot{p_{\mathrm{tot}}}
\def \Anmc{\mathscr{A}_{n}^{m}}
\def \Anmcpw{\mathscr{A}_{n}^{m^{\mathrm{(pw)}}}}
\def \kv{\mathbf{k}}
\def \rv{\mathbf{r}}
\def \rvs{\mathbf{r}_\mathrm{s}}

The process of obtaining the ambisonic signal $\Anmc$, the weights of the spherical basis functions of the three dimensional sound field representing an arbitrary incident field to the microphone array, from the signal captured by the microphone array is referred to as ambisonic \emph{encoding}.

An arbitrary incident field to the spherical microphone array mounted on a rigid sphere with radius $R$ and located at $O$, the origin of the spherical coordinate system $(r,\theta,\varphi)$, can be expanded in terms of the regular spherical basis functions $j_{n}(kr) Y_{n}^{m}(\theta, \varphi)$ of the three-dimensional Helmholtz equation:
\begin{equation}\begin{aligned}
\pin &= \sum_{n=0}^{\infty} \sum_{m=-n}^{n} \Anmc(k) j_{n}(kr) Y_{n}^{m}(\theta, \varphi),
\end{aligned}\label{eq:reconstruction}\end{equation}
with $j_{n}(x)$ the spherical Bessel function of degree $n$ and $k$ the wavenumber.
The total field $\ptot$, which is the sum of the incident field and the scattered field is given by:
\begin{equation}\begin{aligned}
\ptot &= \sum_{n=0}^{\infty} \sum_{m=-n}^{n} \Anmc(k) \left\{ j_n(kr) - h_n(kr) \frac{j_{n}^{\prime}(kR)}{h_{n}^{\prime}(kR)} \right\}  Y_{n}^{m}(\theta, \varphi),\\
\end{aligned}\end{equation}
with $h_n(x)$ the spherical Hankel function of the first kind with degree $n$.
On the surface of the rigid sphere, i.e. $r=R$, this total field is evaluated as:
\begin{equation}\begin{aligned}
\ptot|_{r=R} &= \sum_{n=0}^{\infty} \sum_{m=-n}^{n} \Anmc(k)\frac{ i}{(kR)^{2} h_{n}^{\prime}(kR)} Y_{n}^{m}(\theta, \varphi)\\
\end{aligned}\end{equation}
The total field captured by the $q$-th microphone located at $(R,\theta_q,\varphi_q)$ is therefore given by:
\begin{equation}\begin{aligned}
\ptot^{(q)} &= \sum_{n=0}^{\infty }\sum_{m=-n}^{n}  \frac{ i }{(kR)^{2} h_{n}^{\prime}(kR)}  Y_{n}^{m}(\theta_q, \varphi_q) \Anmc(k).\\
\end{aligned}\end{equation}
By truncating the infinite series with $n_{\mathrm{max}}\equiv N$, this result can be represented in the following vector form:
\begin{equation}\begin{aligned}
\mathbf{p}_{\mathrm{tot}} &= \LambdaB \mathbf{A}\\
\end{aligned}\end{equation}
where $\mathbf{p}_{\mathrm{tot}}$ is a vector holding $\ptot^{(q)}$ in its $q$-th entry (in this paper, indices are 0-based),
$\mathbf{A}$ is a vector holding $\Anmc(k)$ in its $(n^2+n+m)$-th entry,
and $\LambdaB$ is the ``inverse" encoding matrix for rigid sphere microphone arrays which is a matrix holding $\frac{ i }{(kR)^{2} h_{n}^{\prime}(kR)}  Y_{n}^{m}(\theta_q, \varphi_q)$ in its $(q,n^2+n+m)$ entry.
The goal of ambisonic encoding was to obtain $\Anmc(k)$ from the observation $\mathbf{p}_{\mathrm{tot}}$.
Typically, this problem is solved by regularized least squares with a minimization objective:
\begin{equation}\begin{aligned}
L_{\mathrm{enc}} &= |\mathbf{p}_{\mathrm{tot}}-\LambdaB \mathbf{A}|_2 + \sigma |\mathbf{A}|_2,\\
\end{aligned}\end{equation}
with $\sigma$ a regularization parameter, and the solution given by:
\begin{equation}\begin{aligned}
\mathbf{A} &= \argmin_{\mathbf{A}} L_{\mathrm{enc}} = (\LambdaBH \LambdaB + \sigma I)^{-1}\LambdaBH \mathbf{p}_{\mathrm{tot}} = E \mathbf{p}_{\mathrm{tot}},
\end{aligned}\label{eq:sphericalHOAEncoding}\end{equation}
where $E \equiv (\LambdaBH \LambdaB + \sigma I)^{-1}\LambdaBH$ is the regularized encoding matrix.

It is useful to explicitly write down the ambisonic coefficients representing some canonical fields, e.g. plane waves. 
A plane wave with a wave vector in spherical coordinates $(k,\theta_i, \varphi_i)$ is given by:
\begin{equation}\begin{aligned}
\pin^{\mathrm{pw}} = e^{i\kv \cdot \rv}  = \sum _ { n = 0 } ^ { \infty } \sum _ { m = - n} ^ { n } 4 \pi i ^ { n } Y_{n}^{m} \left( \theta _ { i } , \varphi _ { i } \right) ^ { * } j _ { n } ( k r ) Y_{n}^{m} ( \theta , \varphi )
\end{aligned}\end{equation}
Hence, the ambisonic coefficients $\Anmc$ for this plane wave is given by:
\begin{equation}\begin{aligned}
\Anmcpw(k) = 4\pi i^n Y_n^{m}(\theta_i, \varphi_i)^*.
\end{aligned}\end{equation}

\section{Formulation of spheroidal ambisonics}

\def \rl{r_{\mathrm{long}}}
\def \rs{r_{\mathrm{short}}}

The fact that the three-dimensional Helmholtz equation is separable in the spheroidal coordinate system allows us to formulate spheroidal ambisonics.
In this section, the definition of prolate spheroidal coordinates, the definition of spheroidal ambisonic coefficients, the solution of the scattering problem for an arbitrary incoming wave with a rigid prolate spheroid and encoding is presented.

\subsection{Spheroidal coordinates}

While there are two types of spheroidal coordinates, namely the prolate and oblate spheroidal coordinates, 
only the formulation for the case of prolate spheroidal coordinates and prolate spheroidal ambisonics is presented in this paper.
The case of oblate spheroidal coordinates should be able to derive in a similar fashion and could be addressed elsewhere.

The definition of prolate spheroidal coordinates itself has some variations~\cite{flammer2014spheroidal}.
In this paper, the definition also used in \cite{adelman2014software} is employed.
The prolate spheroidal coordinate system has three coordinates $\xi$, $\eta$, and $\varphi$, which is also characterized by the parameter $a$, where
$2a$ is the distance between the two foci of the prolate spheroid. 
The domain of $\xi$ and $\eta$ is $\xi\ge1$ and $|\eta|\le1$, respectively.
The conversion with the Cartesian coordinates $(x,y,z)$ is given by:
\begin{equation}
\left\{
\begin{aligned}
 x =& a  \sqrt{1-\eta^2}  \sqrt{\xi^2 - 1} \cos(\varphi) \\
 y =& a  \sqrt{1-\eta^2}  \sqrt{\xi^2 - 1} \sin(\varphi) \\
 z =& a  \eta \xi
\end{aligned}
\right.
\end{equation}
\begin{equation}
\iff \left\{
\begin{aligned}
\xi =& \frac{1}{2a}  (\sqrt{x^2+y^2+(z+a)^2} + \sqrt{x^2+y^2+(z-a)^2}) \\
\eta =& \frac{1}{2a}  (\sqrt{x^2+y^2+(z+a)^2} - \sqrt{x^2+y^2+(z-a)^2})\\
\varphi =& \arctan(\frac{y}{x})
\end{aligned}
\right.
\end{equation}

The long radius $\rl$ and short radius $\rs$ of a prolate spheroid is related with $a$ and $\xi_1$ by:\\
\begin{equation}
\left\{
\begin{array}{r@{\,}r@{\,}r@{\,}r}
 \rl =& a \xi_{1} \\
 \rs =& a \sqrt{\xi_{1}^{2}-1}
\end{array}
\right.
\end{equation}
\begin{equation}
\iff\left\{
\begin{array}{r@{\,}r@{\,}r@{\,}r}
 a =& \sqrt{\rl^2 - \rs ^2}  \\
 \xi_{1} =& \frac{\rl}{  \sqrt{ \rl^2  - \rs ^2  }   }  = \frac{\rl}{  a} 
\end{array}
\right.
\end{equation}

\subsection{Scattering of an arbitrary incident wave by a sound-hard prolate spheroid}

An arbitrary incident wave can be expanded using radial spheroidal wave functions $R _ { m n } ^ { ( 1 ) }$ and angular spheroidal wave functions  $S _ { m n }$ \cite{flammer2014spheroidal}:
\begin{equation}\begin{aligned}\label{eq_pi_expansion_spheroidal}
\pin = & \sum _ { n = 0 } ^ { \infty } \sum _ { m = 0 } ^ { n } R _ { m n } ^ { ( 1 ) } ( c , \xi ) S _ { m n } ( c , \eta )  \left( A _ { m n } \cos m \varphi + B _ { m n } \sin m \varphi \right)
\end{aligned}\end{equation}
The \emph{spheroidal ambisonic coefficients} are defined as the collection of the $\{A_{mn},B_{mn}\}$ coefficients.
A canonical example of an incident wave is a plane wave $\pin ^ { \mathrm { pw } } = e^{  i \mathbf { k } \cdot \mathbf { r } }$
with a wave vector represented in the Cartesian coordinates: 
\begin{equation}\begin{aligned}
\mathbf { k } = k \left( \sin  \theta _ { 0 }  \cos \varphi_0, \sin \theta_0 \sin \varphi_0 , \cos\theta _ { 0 }  \right),
\end{aligned}\end{equation}
with $k$ the wave number.
The incident plane wave can be expanded as:
\begin{equation}\begin{aligned}
 \pin^{\mathrm{pw}} = &  \sum _ { n = 0 } ^ { \infty } \sum _ { m = 0 } ^ { n } \frac { 2i ^ { n } \varepsilon _ { m } } { N _ { m n } (c)} R _ { m n } ^ { ( 1 ) } ( c , \xi ) S _ { m n } ( c , \eta ) S _ { m n } \left( c , \cos \theta _ { 0 } \right) \cos ( m (\varphi - \varphi_0)  ),
\end{aligned}\end{equation} 
which yields
\begin{equation}
\left\{
\begin{aligned}
 A_{mn}^{\mathrm{pw}} = &   \frac { 2 i ^ { n } \varepsilon _ { m } } { N _ { m n } (c)}  S _ { m n } \left( c , \cos \theta _ { 0 } \right)  \cos m \varphi_0\\
 B_{mn}^{\mathrm{pw}} = &   \frac { 2 i ^ { n } \varepsilon _ { m } } { N _ { m n } (c)}  S _ { m n } \left( c , \cos \theta _ { 0 } \right) \sin m \varphi_0.
\end{aligned}
\right.
\end{equation} 

The total field after scattering an arbitrary incident field characterized by $\{A_{mn},B_{mn}\}$ is then given by:
\begin{equation}\begin{aligned}
\ptot = & \sum _ { n = 0 } ^ { \infty } \sum _ { m = 0 } ^ { n } \left\{ R _ { m n } ^ { ( 1 ) } ( c , \xi ) -  \frac{R_{ m n } ^ { ( 1 )\prime } ( c , \xi_1 )}{ R_{ m n } ^ { ( 3 )\prime } ( c , \xi_1 ) }   R _ { m n } ^ { ( 3 ) } ( c , \xi )   \right\} \\
&\times S _ { m n } ( c , \eta )  \left( A _ { m n } \cos m \varphi + B _ { m n } \sin m \varphi \right).
\end{aligned}\end{equation}

On the surface of the spheroid, i.e. $\xi = \xi_1$, by using the Wronskian relation $W^{(1,3)} = R^{(1)}(c,\xi) R^{(3)\prime}(c,\xi) - R^{(1)\prime}(c,\xi) R^{(3)}(c,\xi) = \frac{i} { c(\xi^2 - 1) } = i W^{(1,2)} $, the total field can be written as:
\begin{equation}\begin{aligned}
\left. \ptot  \right|_{\xi=\xi_1} = & \sum _ { n = 0 } ^ { \infty } \sum _ { m = 0 } ^ { n }  \frac{ i S _ { m n } ( c , \eta )}{ c(\xi_{1}^{2}-1) R_{ m n } ^ { ( 3 )\prime } ( c , \xi_1 ) }    \left( A _ { m n } \cos m \varphi + B _ { m n } \sin m \varphi \right)
\end{aligned}\label{eq:totalFieldOnSurfaceOfProlateSpheroidalScattering}\end{equation}

\subsection{Spheroidal ambisonics encoding}

The goal of spheroidal ambisonic encoding is to estimate the spheroidal ambisonic coefficients from observations by a limited number of microphones mounted on the surface of a spheroid-shaped baffle. As mentioned earlier, it is assumed here that the baffle is a sound-hard prolate spheroid.

By truncating the expansion order by $N>0$, \cref{eq:totalFieldOnSurfaceOfProlateSpheroidalScattering} can be rewritten in vector form:
\begin{equation}\begin{aligned}
\mathbf{p}_{\mathrm{tot}} = (\SA,\SB) \MAT{\RA}{0}{0}{\RB}\VEC{\A}{\B} \equiv  S R \VEC{\A}{\B} \equiv \Lambda^{\mathrm{(P)}} \VEC{\A}{\B}.
\end{aligned}\label{eq:inverseEncodingSpheroidalAmbisonics}\end{equation}
Here, $\mathbf{A}$ and $\mathbf{B}$ are vectors holding $A_{mn}$ and $B_{mn}$ in their $\tilde{l}^{(A)}=((n^2+n)/2+m)$-th and $\tilde{l}^{(B)}=((n^2-n)/2+m-1)$-th entry, respectively.
The lengths of these vectors are $L^{(A)} = \frac{(N+1)(N+2)}{2}$ and $L^{(B)} = \frac{N(N+1)}{2}$, respectively.
The length of the concatenated vector $\VEC{\A}{\B}$ is $L = L^{(A)} + L^{(B)} = (N+1)^2$.
$R$ is a $[L \times L]$ diagonal equalization matrix defined as $R \equiv \MAT{\RA}{0}{0}{\RB}$.
$\RA$ and $\RB$ are diagonal matrices holding $\frac{ i }{ c(\xi_{1}^{2}-1) R_{ m n } ^ { ( 3 )\prime } ( c , \xi_1 ) }$ in their $\tilde{l}^{(A)}$-th and $\tilde{l}^{(B)}$-th diagonal entries, respectively.
$S$ is a concatenation of $\SA$ and $\SB$, with:
\begin{equation}
\left\{
\begin{aligned}
\SA_{q, \tilde{l}^A(m,n)} = S _ { m n } ( c , \eta_q ) \cos m \varphi_q \\ 
\SB_{q, \tilde{l}^B(m,n)} = S _ { m n } ( c , \eta_q ) \sin m \varphi_q \\ 
\end{aligned}
\right.
\end{equation}
where $q$ the sensor index.
$\mathbf{p}_{\mathrm{tot}}$ is a vector holding $\ptot^{(q)}$, the observed sound pressure at the $q$-th microphone, in its $q$-th entry. 
$\mathbf{p}_{\mathrm{tot}}$, $S$, $\SA$, and $\SB$ have shapes of $[Q]$, $[Q \times L]$, $[Q \times L^{(A)}]$, and $[Q \times L^{(B)}]$, respectively, where $Q$ is the number of microphones.
For a truncation order $N$, the total number of unknowns in $\VEC{\A}{\B}$ is $L = (N+1)^2$ unknowns, which is the same as the total number of spherical ambisonics coefficients $\{\Anmc\}$ with maximum order $N$.
$\LambdaP \equiv S R$ is referred to as the ``inverse" encoding matrix for sound-hard prolate spheroidal ambisonics. 

The unknowns $A_{mn}$ and $B_{mn}$ can be estimated from observations of the sound field with multiple sensors mounted on the spheroidal baffle, by solving \cref{eq:inverseEncodingSpheroidalAmbisonics} with least squares.
This process is referred to as \emph{spheroidal ambisonics encoding}.
The regularized least squares solution is given by:
\begin{equation}\begin{aligned}
\VEC{\A}{\B} =  (\LambdaPH \LambdaP + \sigma I)^{-1}\LambdaPH \mathbf{p}_{\mathrm{tot}} = E^{(\mathrm{P})} \mathbf{p}_{\mathrm{tot}},
\end{aligned}\label{eq:spheroidalHOAEncoding}\end{equation}
with $\sigma$ a regularization constant and $E^{(\mathrm{P})} \equiv  (\LambdaPH \LambdaP + \sigma I)^{-1}\LambdaPH $ the encoding matrix for sound-hard prolate spheroidal ambisonics.

\section{Transcoding from spheroidal to spherical ambisonics}

The sound field encoded as a spheroidal ambisonics signal can be converted into a conventional spherical ambisonics representation.
This process is referred to as \emph{transcoding}.
The following relation connecting spheroidal wave functions and spherical Bessel functions and associated Legendre polynomials \cite{flammer2014spheroidal}:
\begin{equation}\begin{aligned}
S _ { m n } ( c , \eta ) R _ { m n } ^ { ( 1) } ( c , \xi ) = \sum _ { r = 0} ^ { \infty } \delta_{(n-m)\%2, r\%2} i^{m-n+r} d _ { r } ^ { m n } ( c ) j _ { m + r } ( k r ) P _ { m + r } ^ { m }  ( \cos \theta ),
\end{aligned}\label{eq:spheroidalToSpherical}\end{equation}
can be utilized for the derivation of the transcoding formula, where $d_r^{mn}(c)$ are the expansion coefficients:
\begin{equation}\begin{aligned}
S _ { m n } ( c , \eta )  = \sum _ { r = 0} ^ { \infty } \delta_{(n-m)\%2, r\%2} d _ { r } ^ { m n } ( c )  P _ { m + r } ^ { m }  ( \eta ).
\end{aligned}\end{equation}
It can be shown that the analytical transcoding formula from spheroidal ambisonics coefficients $\{A_{mn},B_{mn}\}$ to spherical ambisonics coefficients $\mathscr{A}_{n}^{m}$ is given as the following:
\begin{equation}\begin{aligned}
\mathscr{A}_{n^\prime}^{m^{\prime}}
=& I(m^{\prime}) \sqrt{\frac{\pi(n^{\prime}+|m^{\prime}|)! }{(2n^{\prime}+1)(n^{\prime}-|m^{\prime}|)!}} \\ 
&\times  \sum _ { n = |m'| } ^ { \infty }  \delta_{(n-n^\prime)\%2,0} (-1)^{\frac{n'-n}{2}}  d _ { n^\prime-|m^\prime| } ^ { |m^\prime| n } ( c )  \left( A_{|m^\prime| n} - i \mathrm{sgn}(m^{\prime}) B_{|m^\prime| n} \right)  \\
\end{aligned}\label{eq:transcoding}\end{equation}
where
\begin{equation}\begin{aligned}
I(m^{\prime}) = & \begin{cases} 
    (-1)^{m^{\prime}}   & \text{for}\: m^\prime < 0 \\
    2   & \text{for}\: m^\prime = 0 \\
    1   & \text{for}\: m^\prime > 0 \\
\end{cases}\\
\end{aligned}\end{equation}
Here, $\%$ is the modulo operator and $d_{r}^{m,n}$ are the expansion coefficients as defined in \cite{adelman2014software}.

\section{Experimental evaluation}

Prolate spheroidal ambisonic encoding as well as its transcoding into spherical ambisonics was validated by numerical experiments.
Encoding and transcoding of a plane wave with three different incident angles was performed with a sound-hard spherical microphone array as well as a sound-hard prolate spheroidal microphone array.
The spherical array had a radius of 0.198~m. The prolate spheroidal microphone array had $r_{\mathrm{short}}=0.05$~m and $r_{\mathrm{long}}=1$~m. 
The arrays were designed to have the same surface area and both had 512 microphone capsules located on a grid of Gauss-Legendre quadrature nodes for $\theta$ and $\eta$ and equispaced for $\varphi$.
The long axis of the prolate spheroidal array was set parallel to the $x$-axis.
\cref{fig:micArys} shows the experimental procedure and the two microphone arrays used for the experiments.
Spherical and spheroidal ambisonic encoding was performed using \cref{eq:sphericalHOAEncoding} and \cref{eq:spheroidalHOAEncoding}, respectively.
Computation of the coefficient tables of spheroidal wave functions were performed using the software library \emph{Spheroidal}~\cite{adelman2014software}.
The truncation order was set to $N=12$ for both spherical and spheroidal ambisonics.
The regularization parameter $\sigma$ was set to zero for both spherical and spheroidal encoding, i.e. no regularization was applied.
Transcoding from spheroidal ambisonics to spherical ambisonics was performed using \cref{eq:transcoding}, truncated for $n\le N$. 
The estimated incident field for the encoded spherical ambisonic coefficients was reconstructed and compared to the ground truth incident field.
The reconstruction of the estimated incident fields was performed using \cref{eq:reconstruction} truncated for $n\le N$.
The signal-to-distortion ratio (SDR) of the reconstructed fields was computed for evaluation points in the $x-y$ plane. 
The region with SDR higher than 30 dB was considered as the sweet-spot of accurate reconstruction.

\begin{figure}[htbp]
\centering
\includegraphics[width=12cm]{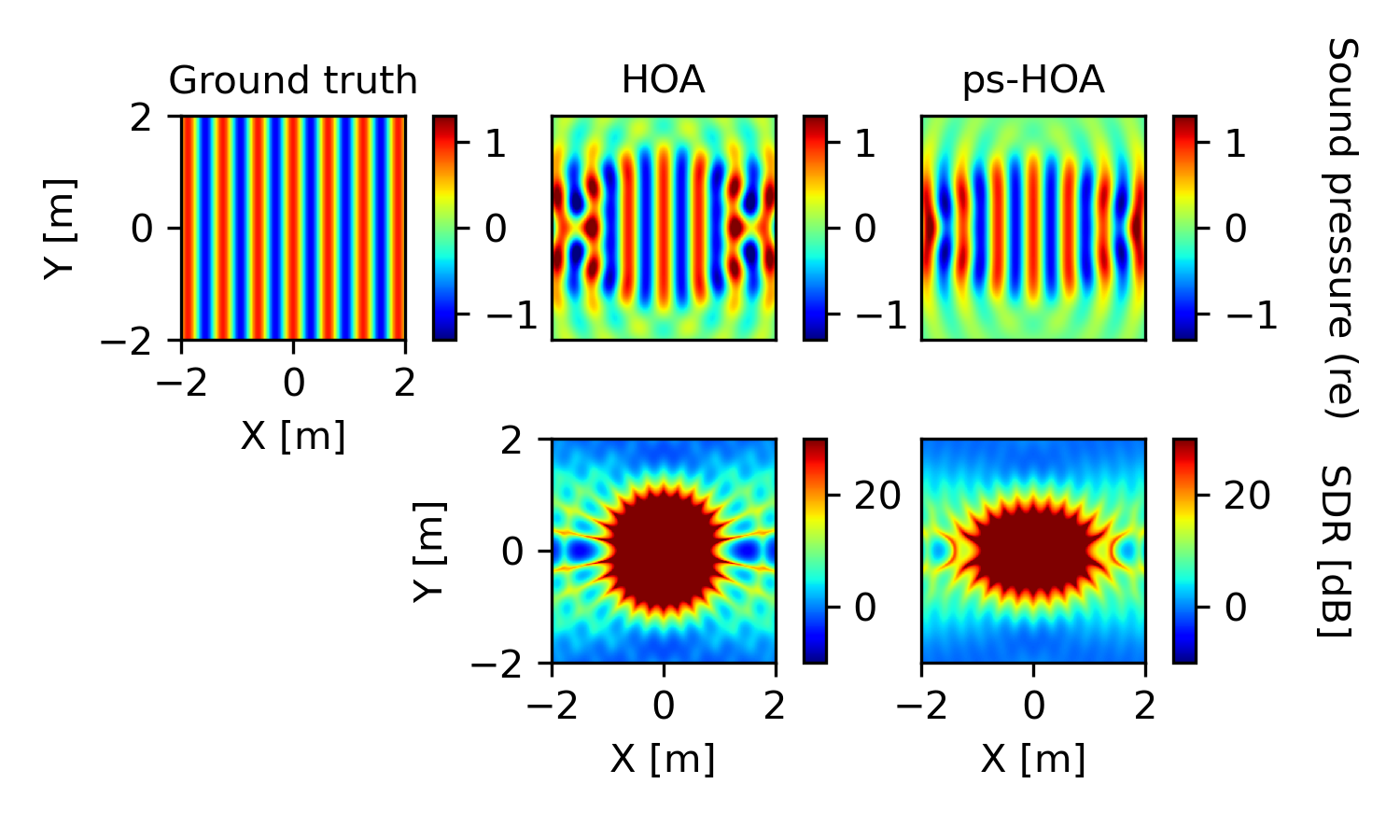}
\caption{Results for an incident plane wave travelling along the long axis of the spheroidal array, which is set parallel to the $x$-axis. The first row from left to right: the real part of the sound pressure of the ground truth incident field, the field reconstructed from spherical ambisonic (HOA) coefficients, and the field reconstructed from the prolate spheroidal ambisonic (ps-HOA) coefficients transcoded to spherical ambisonic coefficients. The second row presents the SDR of the reconstructed fields for HOA (left) and ps-HOA (right). The region with SDR higher than 30 dB was considered as the sweet-spot and is colored in red.}
\label{fig:fields_x}
\end{figure}

\begin{figure}[htbp]
\centering
\includegraphics[width=12cm]{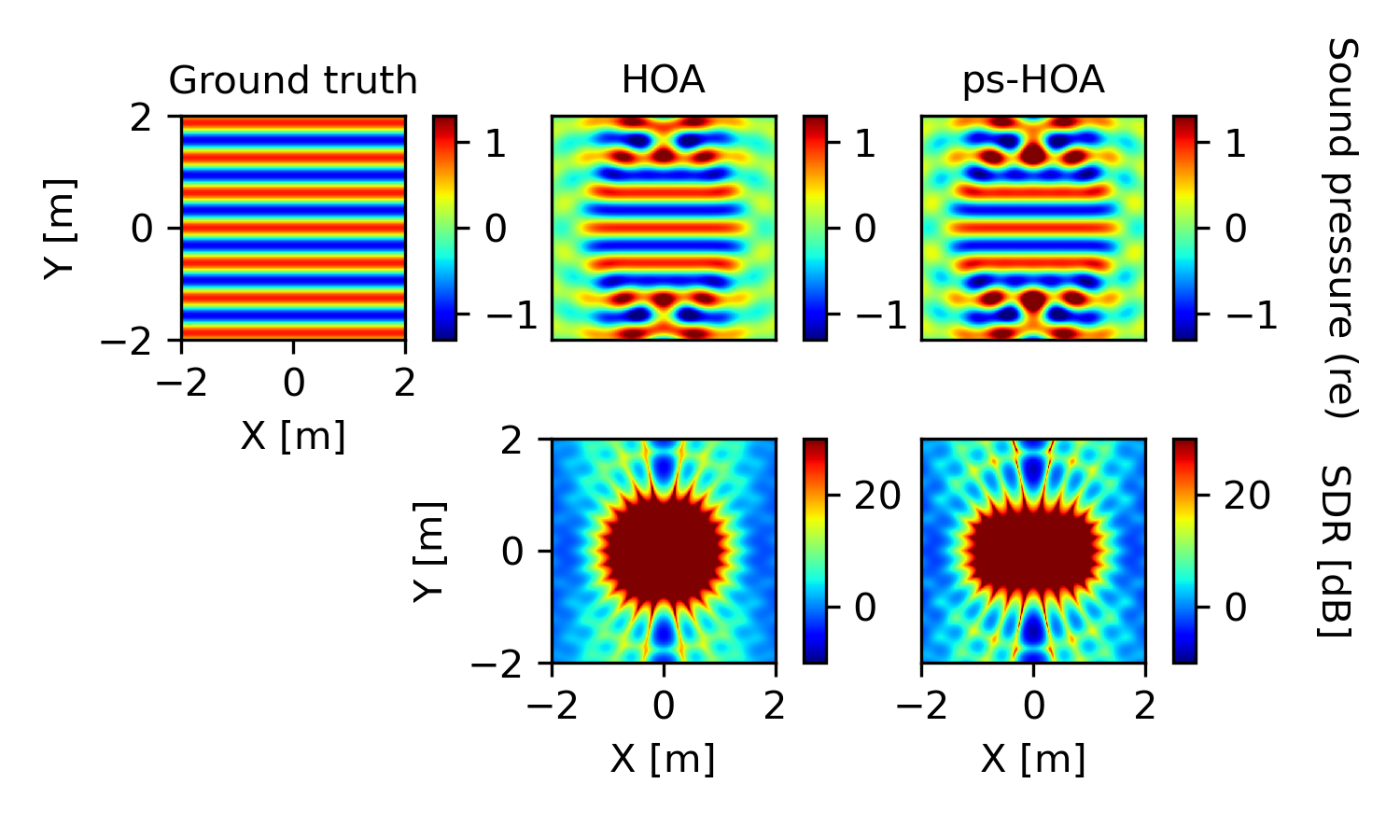}
\caption{Results for an incident plane wave travelling along the short axis of the spheroidal array which is set to the $y$-axis. The definition of each of the subplot is identical to \cref{fig:fields_x}.}
\label{fig:fields_y}
\end{figure}

\begin{figure}[htbp]
\centering
\includegraphics[width=12cm]{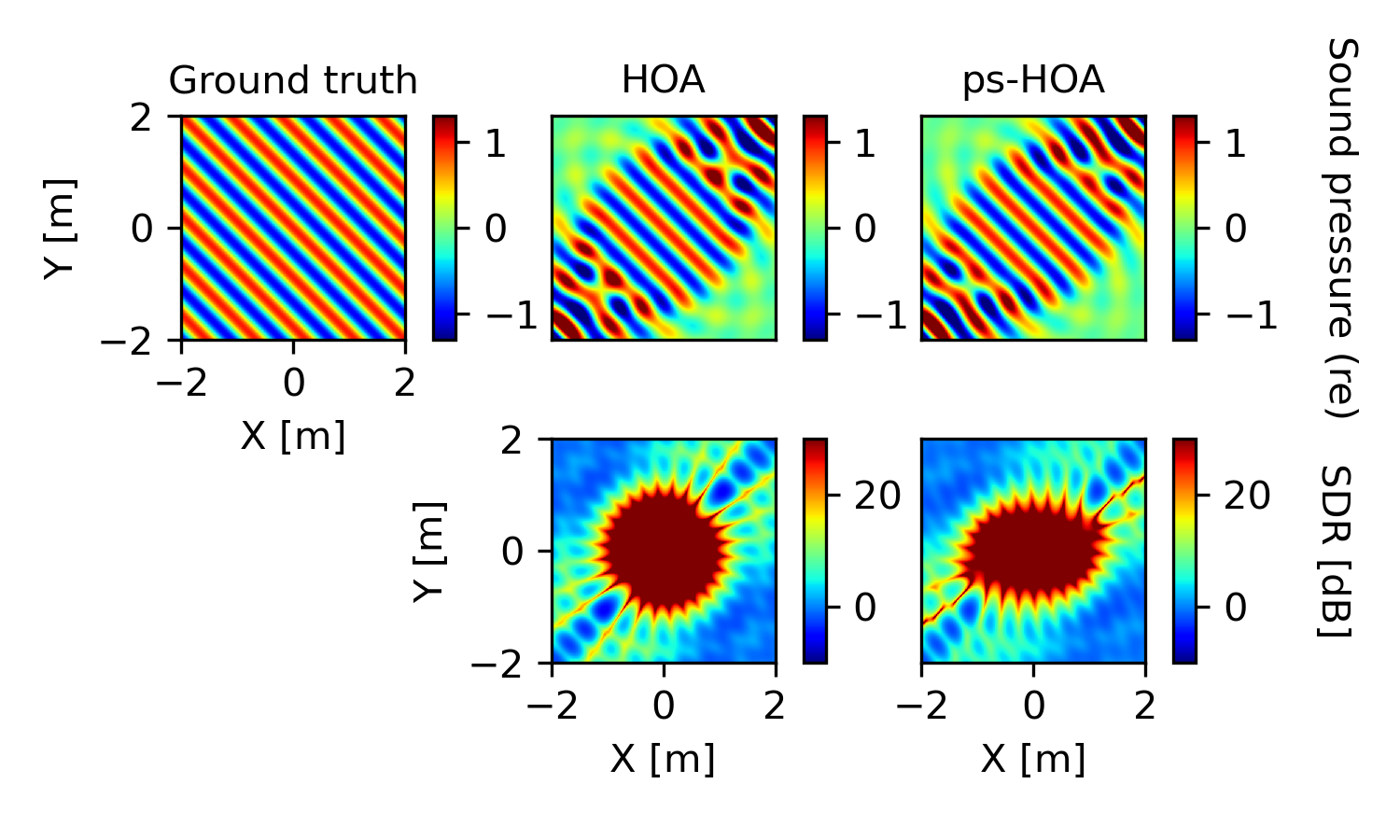}
\caption{Results for an incident plane wave with a normalized wave vector $(\frac{\sqrt{2}}{2},\frac{\sqrt{2}}{2},0)$. The definition of each of the subplot is identical to \cref{fig:fields_x}.}
\label{fig:fields_xy}
\end{figure}

\cref{fig:fields_x}, \cref{fig:fields_y}, and \cref{fig:fields_xy} shows the results for incident waves with normalized wave vectors, expressed in the Cartesian coordinates, of $(1,0,0)$, $(0,1,0)$, and $(\frac{\sqrt{2}}{2},\frac{\sqrt{2}}{2},0)$, respectively.
The frequency of the incident wave was 541.8~Hz.
It can be observed that the width of the sweet-spot of precise reconstruction in spheroidal ambisonics is shorter in the shorter axis of the spheroid, but longer in the longer axis of the spheroid, compared to the width in the baseline spherical ambisonics case.
This asymmetry of the sweet-spot shape could be useful in some applications, in which a non-spherical sweet-spot is desired.
An example application is sound field reproduction for multi-person home-theater systems in which the sweet-spot should cover multiple listeners sitting next to each other.

\section{Conclusion}
The framework of spheroidal ambisonics, a natural extension of ambisonics into spheroidal coordinates, was proposed.
Spheroidal ambisonics enables analytical encoding of the spatial sound field into spheroidal ambisonic coefficients using spheroidal microphone arrays.
An analytical transcoding formula from spheroidal ambisonics into conventional spherical ambisonics was derived, in order to ensure compatibility with the existing software ecosystem around spherical ambisonics.
The numerical experiments demonstrated that the sweet-spot of reconstruction in spheroidal ambisonics has an asymmetric shape which is prolonged towards the longer axis of the prolate spheroidal microphone array, realizing non-spherical sweet-spots in ambisonic reconstruction, which could be useful in some applications.
The case of oblate spheroidal microphone arrays can be derived in a similar fashion and will be published elsewhere.
A recently proposed microphone array for three-dimensional ambisonics recording, which uses a sound-hard circular disc as the scattering body~\cite{berge2019acoustically}, can be seen as a special case of an oblate spheroidal ambisonic microphone array.
Another future research topic is the optimization of the microphone capsule configuration on the spheroid.
In a practical setup, care must be taken for spatial aliasing~\cite{rafaely2007spatial} and a careful design of the microphone array configuration is important.
While the subject of optimizing the microphone array configuration for spherical arrays has been studied extensively in the past~\cite{li2004flexible, li2007flexible}, optimization of the array configuration in the case of spheroidal microphone arrays requires further research.

\section*{Acknowledgments}
The author thanks Professor Dr. Ramani Duraiswami at the University of Maryland, College Park for providing feedback for the manuscript.

\bibliographystyle{unsrt}
\bibliography{main}

\end{document}